\documentclass{appolb}
\usepackage{graphicx}
\usepackage[colorlinks=true,linktocpage=true,linkcolor=blue,citecolor=blue]{hyperref}
\usepackage{cite}
\usepackage[nice]{nicefrac}
\def\HP{\hphantom{\alpha}} % horizontal

\def\be{\begin{equation}}
	\def\ee{\end{equation}}
\newcommand{\bel}[1]{\begin{eqnarray}\label{#1}}
	\newcommand{\eel}{\end{eqnarray}}
\def\barr{\begin{array}}
	\def\earr{\end{array}}
\def\beq{\begin{eqnarray}}
	\def\eeq{\end{eqnarray}}
\def\bfig{\begin{figure}}
	\def\efig{\end{figure}}

\def\CHI{\chi}
\newcommand{\nn}{\nonumber}
\newcommand{\f}[2]{\frac{#1}{#2}}
\newcommand{\onehalf}{{\nicefrac{1}{2}}}

\def\a{\alpha}
\def\b{\beta}
\def\g{\gamma}
\def\d{\delta}

\def\LR{\left(} % round
\def\RR{\right)}
\def\HP{\hphantom{\alpha}} % horizontal

\newcommand{\sh}[1]{\sinh#1}
\newcommand{\ch}[1]{\cosh#1}

\def\half{\frac{1}{2}}

\def\GLW{{\rm GLW}}

\newcommand{\lab}[1]{\label{#1}}
\def\nn{\nonumber}

\def\half{\frac{1}{2}}

\def\GLW{{\rm GLW}}

\def\n0{n_{(0)}}
\def\e0{\varepsilon_{(0)}}
\def\P0{P_{(0)}}
\begin{document}
\title{Boost-invariant description of polarization within hydrodynamics with spin%
\thanks{Presented at 45. Zjazd Fizyk\'ow Polskich PTF 2019, September 13-18, 2019, Krak\'ow.}%
}
\author{Rajeev Singh
\address{Institute of Nuclear Physics Polish Academy of Sciences, PL-31-342 Krak\'ow, Poland}
% {Third Author of different affiliation
% }
% the Name(s) of other Author(s)
% \address{affiliation}
}
\maketitle
\begin{abstract}
We briefly review a recently proposed formalism of perfect-fluid hydrodynamics with spin, which is a generalization of the standard hydrodynamic framework and provides a natural tool for describing the evolution of spin-polarized systems of particles with spin 1/2. It is based on the de Groot - van Leeuwen - van Weert forms of energy-momentum and spin tensors and conservation laws. Using Bjorken symmetry we show how this formalism may be used to determine  observables describing the polarization of particles  measured in the experiment.
\end{abstract}
  
\section{Introduction} 
The spin polarization measurements of $\Lambda$ hyperons recently made by the STAR Collaboration~\cite{STAR:2017ckg,Adam:2018ivw,Adam:2019srw,Niida:2018hfw} prompted vast theoretical developments aiming at understanding the relation between the orbital angular momentum of the matter created in relativistic heavy-ion collisions and the average spin orientation of the particles emitted from such systems \cite{Becattini:2009wh,Becattini:2013fla,Montenegro:2017rbu,Becattini:2018duy,Boldizsar:2018akg,
 Prokhorov:2018bql,Yang:2018lew,Florkowski:2019voj,Weickgenannt:2019dks,Hattori:2019lfp,Ambrus:2019ayb,Sheng:2019kmk,Prokhorov:2019cik,Ivanov:2019wzg,Hattori:2019ahi,Xie:2019jun,Liu:2019krs,Wu:2019eyi,Becattini:2019ntv,Zhang:2019xya,Li:2019qkf,Florkowski:2019gio,Deng:2020ygd,Fukushima:2020qta}. Lately it has been seen that the thermal-based models which successfully describe the global spin polarization~\cite{Becattini:2016gvu,Karpenko:2016jyx,Li:2017slc,Xie:2017upb}, unfortunately fail at explaining differential results \cite{Adam:2019srw}. These models assume that the spin polarization at the freeze-out is entirely determined by the so called thermal vorticity~\cite{Becattini:2007sr,Becattini:2013fla} and lack the dynamical evolution of the spin polarization which takes place in the system's expansion. Following the ideas of Refs.~\cite{Florkowski:2017ruc,Florkowski:2017dyn,Florkowski:2018fap,Florkowski:2018ahw}, we investigate this possibility by extending the standard perfect-fluid hydrodynamic framework to include the dynamics of the spin degrees of freedom and analysing it in the Bjorken symmetry setup~\cite{Florkowski:2019qdp}. 

\section{Hydrodynamic equations}
 
 The perfect-fluid hydrodynamics for spin-$\onehalf$ particles is constructed  based on the conservation laws for charge, energy, linear momentum and angular momentum with the de Groot - van Leeuwen - van Weert (GLW)~\cite{DeGroot:1980dk}  forms of the energy-momentum tensor, $T^{\a\b}_{\rm GLW}$, and spin  tensor, $S^{\a\b\g}_{\rm GLW}$\footnote{Herein we assume that spin polarization is small ($|\omega_{\mu\nu}| < 1$).}, namely~\cite{Florkowski:2017ruc,Florkowski:2017dyn,Florkowski:2018fap}
\begin{eqnarray} 
\quad\partial_\mu N^\mu = 0,  \qquad
\partial_\mu T^{\mu\nu}_{\rm GLW} = 0, \qquad
\partial_\lambda  S_{\rm GLW}^{\lambda, \alpha \beta} =T_{\rm GLW}^{\beta \alpha} -T_{\rm GLW}^{\alpha \beta}, 
\label{eom}
\end{eqnarray}
with
\vspace{-0.3cm}
\bel{Tmn}
N^\alpha = n U^\alpha, \qquad T^{\a\b}_{\rm GLW} = (\varepsilon + P ) U^\a U^\b - P g^{\a\b},
\eel
where $N^\alpha$ is the net baryon charge current, $\varepsilon$ is the energy density, $P$ is the pressure, $n$ is the baryon density and $U^\beta$ is the time-like fluid flow four-vector. Since GLW energy-momentum tensor is symmetric in Eq.~(\ref{eom}) the angular momentum conservation implies separate conservation of the spin part~\cite{Florkowski:2018ahw}. The spin current is given by $S^{\alpha , \beta \gamma }_{\rm GLW}
=  {\cal C} \left( n_{(0)}(T) U^\alpha \omega^{\beta\gamma}  +  S^{\a, \b\g}_{\Delta\GLW} \right)$, where ${\cal C} = \ch(\xi)$ and the  auxiliary tensor $S^{\a, \b\g}_{\Delta\GLW}$ is defined as~\cite{Florkowski:2017dyn}
\beq
S^{\a, \b\g}_{\Delta\GLW} 
&=&  {\cal A}_{(0)} \, U^\a U^\d U^{[\b} \omega^{\g]}_{\HP\d} \lab{SDeltaGLW} \\
&& \hspace{-0.5cm} + \, {\cal B}_{(0)} \, \Big( 
U^{[\b} \Delta^{\a\d} \omega^{\g]}_{\HP\d}
+ U^\a \Delta^{\d[\b} \omega^{\g]}_{\HP\d}
+ U^\d \Delta^{\a[\b} \omega^{\g]}_{\HP\d}\Big),
\nn
\eeq
with $
{\cal B}_{(0)} =-\frac{2}{\hat{m}^2} s_{(0)}(T)$ and $ 
{\cal A}_{(0)}  = -3{\cal B}_{(0)} +2 n_{(0)}(T)$, 
where $n_{(0)}(T)$ and $s_{(0)}(T)$ are the number density and entropy density of spin-less and neutral massive Boltzmann particles,  $T$ is the temperature, $\Delta^{\a\b}$ is the projector on the spatial direction to $U$, $\xi$ is the ratio of baryon chemical potential, $\mu$, and temperature, $T$, and  $\hat{m}$ is the ratio of the particle mass and temperature.

\section{Spin polarization tensor and boost-invariant flow}

The polarization tensor $\omega_{\mu\nu}$ can be decomposed in the following way
\beq
\omega_{\mu\nu} &=& \kappa_\mu U_\nu - \kappa_\nu U_\mu + \epsilon_{\mu\nu\a\b} U^\a \omega^{\b}, \lab{spinpol1}
\eeq
where $\kappa$ and $\omega$ are four-vectors orthogonal to $U$. 
For boost-invariant and transversely homogeneous systems we introduce the following basis
\beq
U^\a &=& \frac{1}{\tau}\LR t,0,0,z \RR = \LR \ch(\eta), 0,0, \sh(\eta) \RR, \nn \\
X^\a &=& \LR 0, 1,0, 0 \RR,\nn\\
Y^\a &=& \LR 0, 0,1, 0 \RR, \nn\\
Z^\a &=& \frac{1}{\tau}\LR z,0,0,t \RR = \LR \sh(\eta), 0,0, \ch(\eta) \RR, 
\lab{BIbasis}
\eeq
where
$\tau = \sqrt{t^2-z^2}$ is the longitudinal proper time and $\eta = \half \ln((t+z)/(t-z))$ is the space-time rapidity.\\
Using above basis, one can decompose the vectors $\kappa^{\mu}$ and $\omega^{\mu}$ as
\beq
\kappa^\a &=&  C_{\kappa X} X^\a + C_{\kappa Y} Y^\a + C_{\kappa Z} Z^\a, \lab{eq:k_decom}\\
\omega^\a &=&  C_{\omega X} X^\a + C_{\omega Y} Y^\a + C_{\omega Z} Z^\a, \lab{eq:o_decom}
\eeq
where the coefficients ${C}_{\kappa X}$, 
${C}_{\kappa Y}$, ${C}_{\kappa Z}$, ${C}_{\omega X}$, ${C}_{\omega Y}$, and ${C}_{\omega Z}$ are functions of $\tau$ only.  
Using the above forms of $\kappa^\alpha$ and $\omega^\alpha$ in conservation law of spin tensor and projecting the resulting tensor equation on $U_\mu X_\nu$, $U_\mu Y_\nu$, $U_\mu Z_\nu$, $X_\mu Y_\nu$, $X_\mu Z_\nu$ and $Y_\mu Z_\nu$, we obtain the set of the equations for the coefficients $C$. These coefficients turn out to evolve independently. The scalar functions ${C}_{\kappa X}$ and ${C}_{\kappa Y}$ (and similarly ${C}_{\omega X}$ and ${C}_{\omega Y}$) obey the same form of differential equations due to the rotational invariance in the transverse plane. 
%%%%%%%%%%%%%%
\section{Information about spin polarization of particles at freeze-out}
The knowledge about the evolution of spin polarization tensor allows us to calculate the average spin polarization per particle which is defined by $\langle\pi_{\mu}\rangle=E_p\frac{d\Pi _{\mu }(p)}{d^3 p}/E_p\frac{d{\cal{N}}(p)}{d^3 p}$~\cite{Florkowski:2018ahw} with
\beq
E_p\frac{d\Pi _{\mu }(p)}{d^3 p} = -\f{ \cosh(\xi)}{(2 \pi )^3 m}
\int
\Delta \Sigma _{\lambda } p^{\lambda } \,
e^{-\beta \cdot p} \,
\tilde{\omega }_{\mu \beta }p^{\beta }, 
\eeq
and 
\beq
E_p\frac{d{\cal{N}}(p)}{d^3 p}&=&
\f{4 \cosh(\xi)}{(2 \pi )^3}
\int
\Delta \Sigma _{\lambda } p^{\lambda } 
\,
e^{-\beta \cdot p},
\eeq
where $E_p\frac{d\Pi _{\mu }(p)}{d^3 p}$ is the total value (integrated over freeze out hypersuface $\Delta \Sigma_{\lambda }$) of the Pauli-Luba\'nski vector for particles with momentum $p$ and $E_p\frac{d{\cal{N}}(p)}{d^3 p}$ is the momentum density of all particles.\\
In the particle rest rame (PRF), using canonical boost one can get the polarization vector $\langle\pi^{\star}_{\mu}\rangle$. Its longitudinal component is given as~\cite{Florkowski:2019qdp}
 \begin{eqnarray}
\langle\pi^{\star}_{z}\rangle&=& \frac{1}{8m}
\Big[\left(\frac{m\ch(y_p)+m_T}{m_T \ch(y_p)+m}\right)\left[\CHI\left(C_{\kappa X} p_y-C_{\kappa Y} p_x\right)+2 C_{\omega Z} m_T  \right]\nn \\
&& +\frac{\CHI \,m\,\sh(y_p) \left(C_{\omega X} p_x+C_{\omega Y} p_y\right)}{m_T \ch(y_p)+m}\Big],
 \end{eqnarray}
where $\CHI\left( \hat{m}_T \right)=\left( K_{0}\left( \hat{m}_T \right)+K_{2}\left( \hat{m}_T \right)\right)/K_{1}\left( \hat{m}_T \right)$, $\hat{m}_T$ is the ratio of transverse mass ($m_T$), and temperature ($T$) and $y_p$ is the rapidity.

\section{Numerical Results}
\begin{figure}
\centering
 \includegraphics[width=0.49\textwidth]{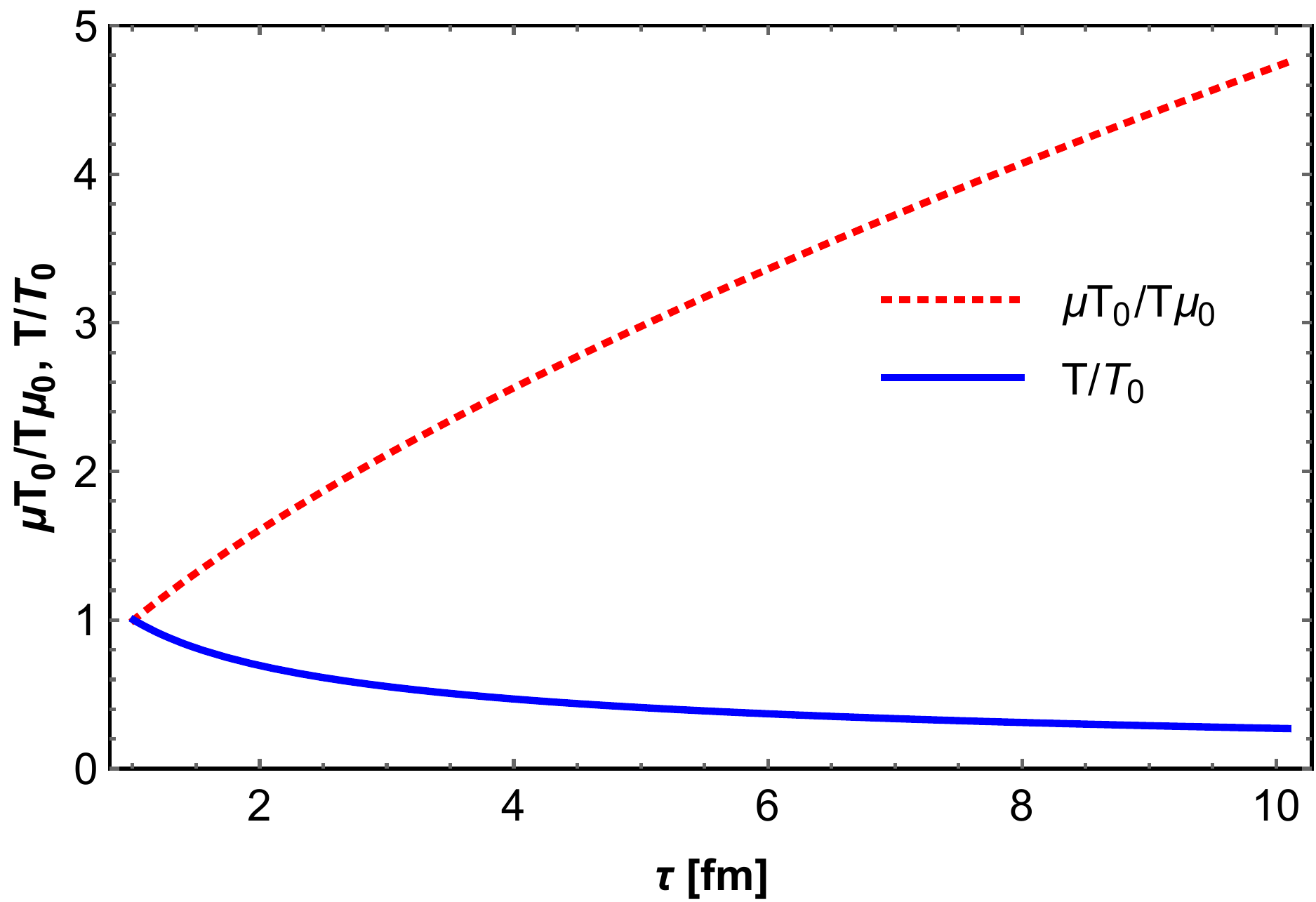}
 \includegraphics[width=0.49\textwidth]{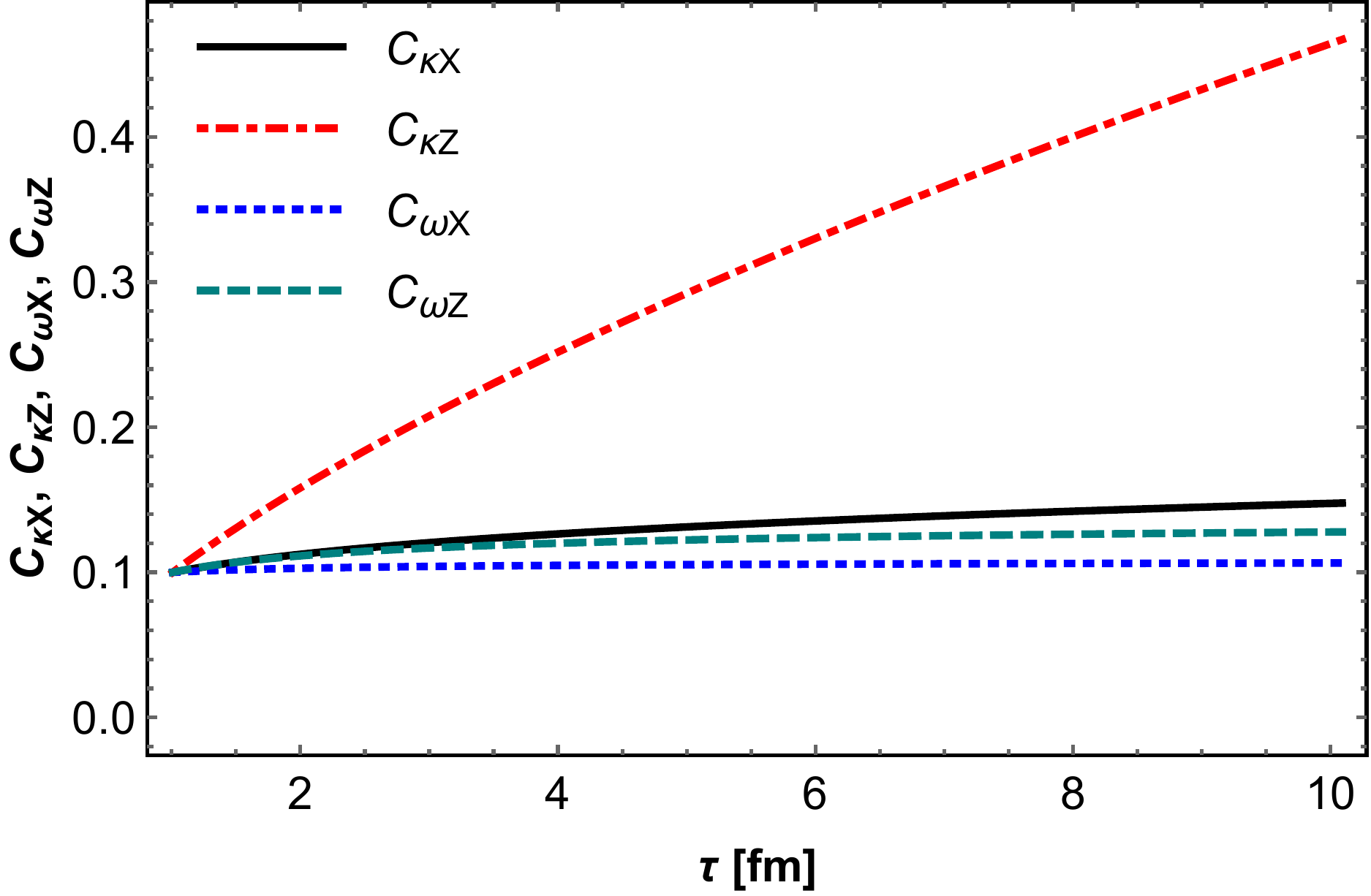}
\caption{Left: Proper-time dependence of temperature $T$ divided by its initial value $T_0$ (solid line) and the ratio of $\mu$ (baryon chemical potential) and $T$ (temperature) rescaled by the initial ratio $\mu_0/T_0$ (dotted). Right: Proper-time dependence of the coefficients $C_{\kappa X}$, $C_{\kappa Z}$, $C_{\omega X}$ and $C_{\omega Z}$.}
\label{fig:FH}
\end{figure}

Here, we present the numerical solutions of boost-invariant forms of the conservation laws. For the Bjorken geometry, conservation of charge current can be written as $\frac{dn}{d\tau}+\frac{n}{\tau}=0$ and conservation of energy and linear momentum can be written as $\frac{d\varepsilon}{d\tau}+\frac{(\varepsilon+P)}{\tau}=0$.
In Fig.\ref{fig:FH} (left), we show the proper-time dependence of temperature and baryon chemical potential obtained from boost-invariant forms of conservation laws. We have reproduced the established results that the temperature decreases with proper-time and ratio of chemical potential over temperature increases with proper time. In Fig.\ref{fig:FH} (right), we show the proper time dependence of the $C$ coefficients that define the spin polarization.

\begin{figure}
\centering
 \includegraphics[width=0.32\textwidth]{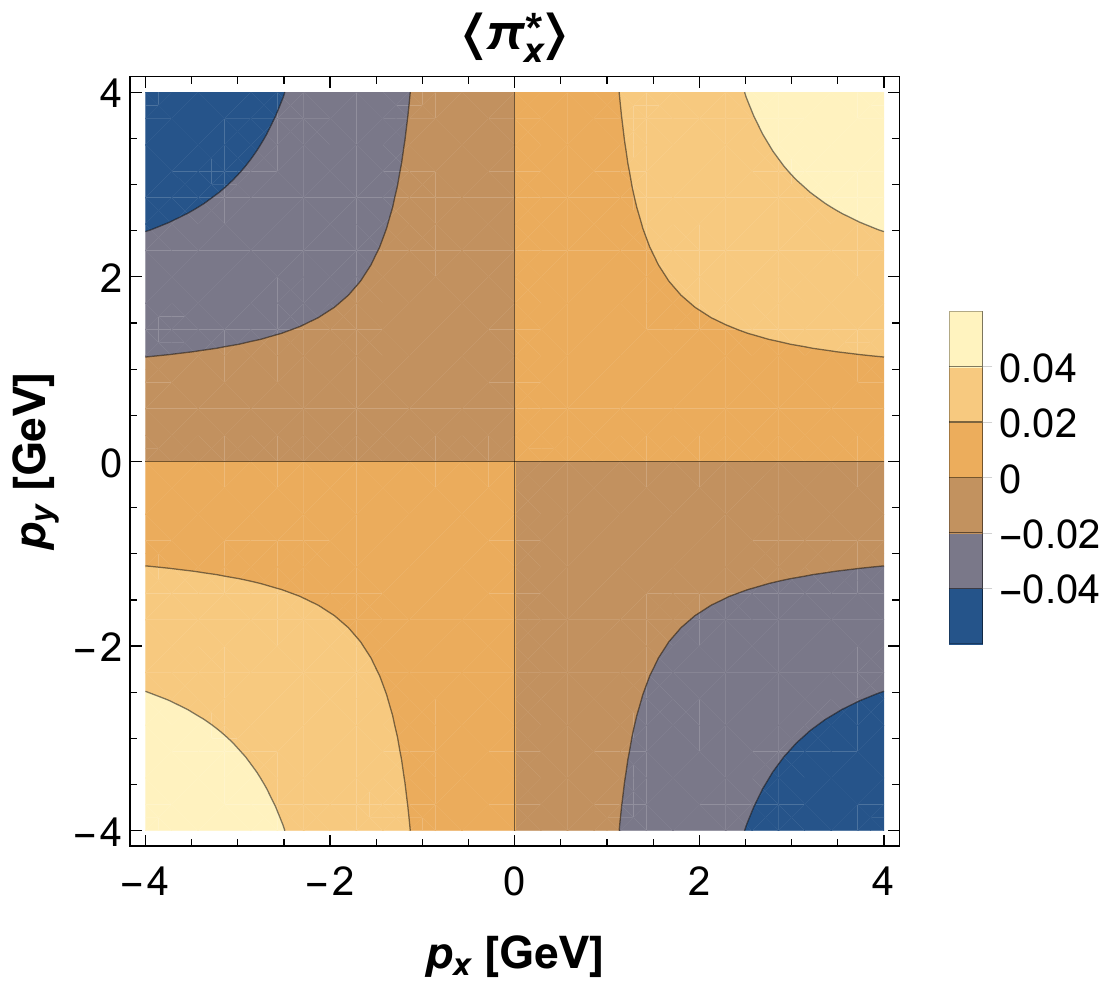}
 \includegraphics[width=0.32\textwidth]{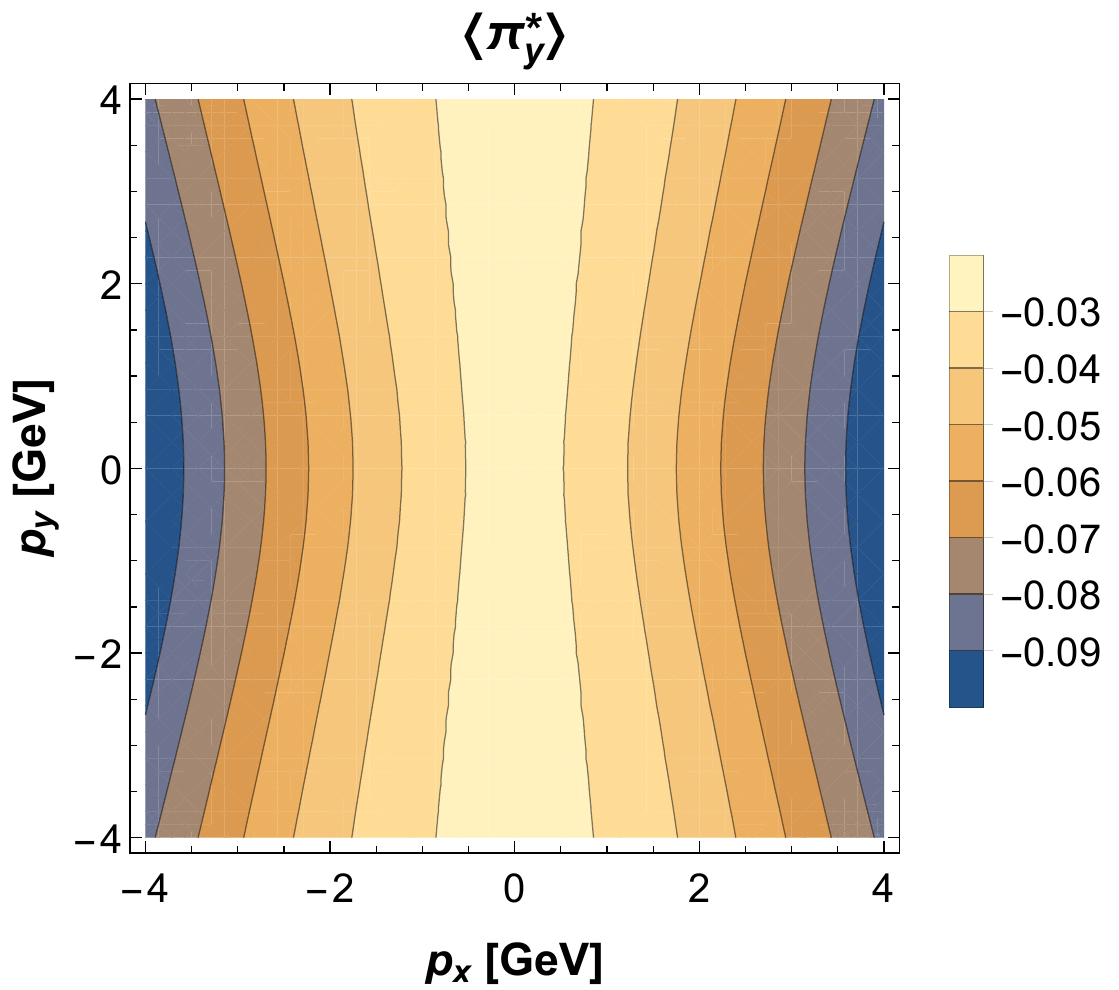}
 \includegraphics[width=0.32\textwidth]{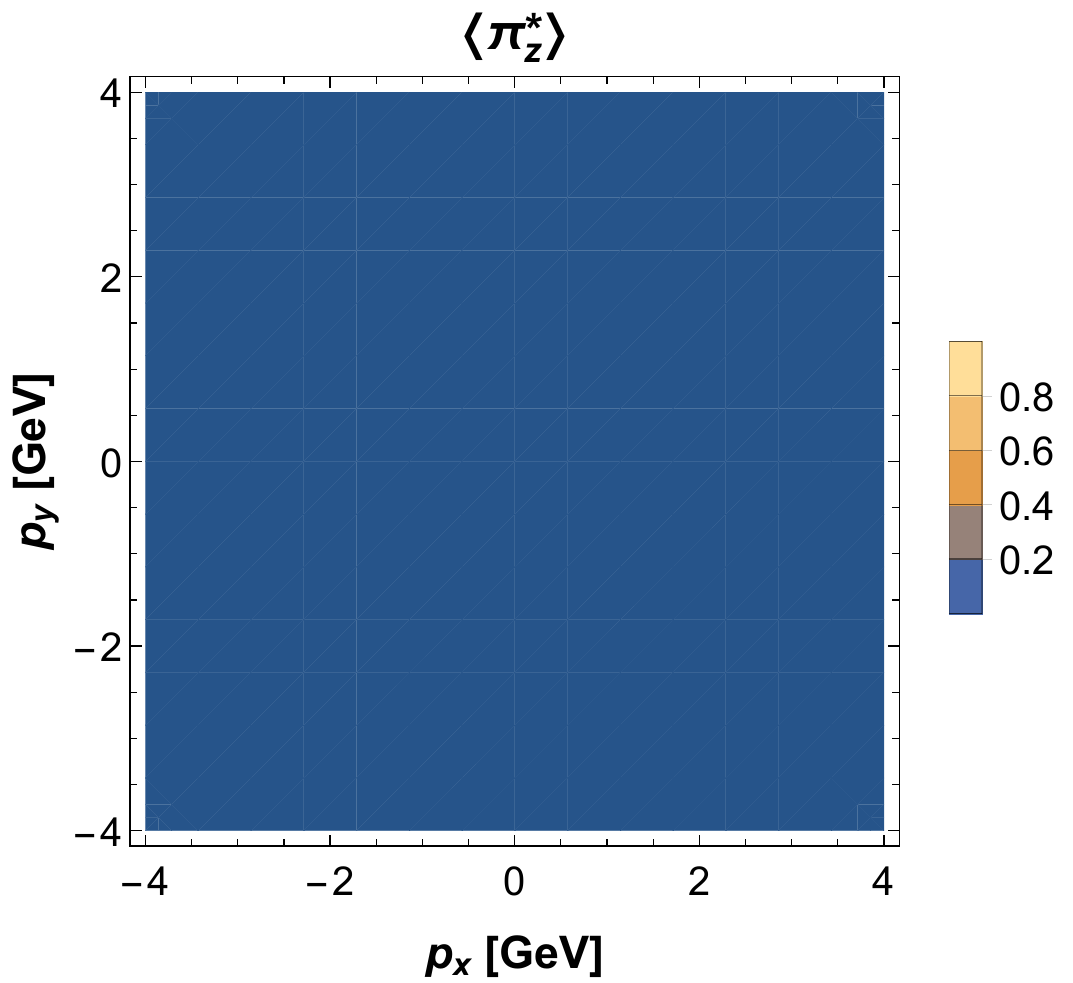}
\caption{Components of the PRF mean polarization three-vector of $\Lambda$'s with the initial conditions $\mu_0=800$~MeV,
$T_0=155$~MeV, $C_{\kappa, 0}=(0,0,0)$ and $ C_{\omega, 0}=(0,0.1,0)$ for $y_p=0$.}
\label{fig:polarization1}
\end{figure}

Using the values of thermodynamic parameters and $C$ coefficients at freeze-out, we can get the different components of the PRF  mean polarization  vector $\langle\pi^{\star}_{\mu}\rangle$ as the functions of particle three-momentum, see Fig.\ref{fig:polarization1}.
We observe that  $\langle\pi^{\star}_{y}\rangle$ is negative, reflecting the initial spin angular momentum of the system (the original collision process has only orbital angular momentum perpendicular to the reaction plane and its direction is opposite to the y axis). Since the experiments are done at midrapidity, the longitudinal component ($\langle\pi^{\star}_{z}\rangle$) is zero and $\langle\pi^{\star}_{x}\rangle$ shows quadrupole structure.
These results does not reproduce the observed experimental quadrupole structure of the longitudinal polarization because of the symmetries we have in our model.
%%%%%%
\newpage
\section{Conclusion}
Using the perfect-fluid hydrodynamics with spin we have presented the numerical results describing the space-time evolution of the spin polarization tensor for a Bjorken hydrodynamic background~\cite{Bjorken:1982qr}. Our approach is based on the GLW forms of the energy-momentum and spin tensors assuming the spin polarization tensor in the leading order.  It has been shown that six scalar functions $C$ describing spin polarization evolve independently of each other and their proper-time dependence is weak. These results can also be used for the determination of the spin polarization of particles at the freeze-out hypersurface. We have also shown that the spin polarization of particles formed at freeze out reflects the initial direction of polarization. 
\medskip

This research was supported in part by the Polish National Science Center Grants No. 2016/23/B/ST2/00717 and No. 2018/30/E/ST2/00432.


\begin{thebibliography}{99}
 
 %\cite{STAR:2017ckg}
\bibitem{STAR:2017ckg} 
  L.~Adamczyk {\it et al.} [STAR Collaboration],
  %``Global $\Lambda$ hyperon polarization in nuclear collisions: evidence for the most vortical fluid,''
  \textcolor{red}{\textit{Nature} {\bf 548}, 62 (2017)}.
 % doi:10.1038/nature23004
%  [arXiv:1701.06657 [nucl-ex]].

%\cite{Adam:2018ivw}
\bibitem{Adam:2018ivw} 
  J.~Adam {\it et al.} [STAR Collaboration],
  %``Global polarization of $\Lambda$ hyperons in Au+Au collisions at $\sqrt{s_{_{NN}}}$ = 200 GeV,''
  \textcolor{red}{\textit{Phys.\ Rev.\ C} {\bf 98}, 014910 (2018)}.
 % doi:10.1103/PhysRevC.98.014910
 % [arXiv:1805.04400 [nucl-ex]].

%\cite{Niida:2018hfw}
\bibitem{Niida:2018hfw} 
  T.~Niida [STAR Collaboration],
  %``Global and local polarization of $\Lambda$ hyperons in Au+Au collisions at 200 GeV from STAR,''
  \textcolor{red}{\textit{Nucl.\ Phys.\ A} {\bf 982}, 511 (2019)}.
  %doi:10.1016/j.nuclphysa.2018.08.034
 % [arXiv:1808.10482 [nucl-ex]].

 %\cite{Adam:2019srw}
\bibitem{Adam:2019srw} 
  J.~Adam {\it et al.} [STAR Collaboration],
  %``Polarization of $\Lambda$ ($\bar{\Lambda}$) hyperons along the beam direction in Au+Au collisions at $\sqrt{s_{_{NN}}}$ = 200 GeV,''
 \textcolor{red}{\textit{Phys.\ Rev.\ Lett.\ } {\bf 123}, no. 13, 132301 (2019)}.
 % doi:10.1103/PhysRevLett.123.132301
 % [arXiv:1905.11917 [nucl-ex]].
  %%CITATION = doi:10.1103/PhysRevLett.123.132301;%%
  %10 citations counted in INSPIRE as of 15 Jan 2020
  
%\cite{Becattini:2013fla}
\bibitem{Becattini:2013fla} 
  F.~Becattini \textit{et al.},
  %``Relativistic distribution function for particles with spin at local thermodynamical equilibrium,''
  \textcolor{red}{\textit{Annals Phys.\ } {\bf 338}, 32 (2013)}.
 % doi:10.1016/j.aop.2013.07.004
%  [arXiv:1303.3431 [nucl-th]].
  
  %\cite{Becattini:2018duy}
\bibitem{Becattini:2018duy} 
  F.~Becattini, W.~Florkowski and E.~Speranza,
  %``Spin tensor and its role in non-equilibrium thermodynamics,''
  \textcolor{red}{\textit{Phys.\ Lett.\ B} {\bf 789}, 419 (2019)}.
 % doi:10.1016/j.physletb.2018.12.016
 % [arXiv:1807.10994 [hep-th]].
 
  %\cite{Becattini:2009wh}
\bibitem{Becattini:2009wh} 
  F.~Becattini and L.~Tinti,
  %``The Ideal relativistic rotating gas as a perfect fluid with spin,''
  \textcolor{red}{\textit{Annals Phys.\ } {\bf 325}, 1566 (2010)}.
 % doi:10.1016/j.aop.2010.03.007
 % [arXiv:0911.0864 [gr-qc]].
   
 
  %\cite{Montenegro:2017rbu}
\bibitem{Montenegro:2017rbu} 
  D.~Montenegro \textit{et al.},
  %``Ideal relativistic fluid limit for a medium with polarization,''
  \textcolor{red}{\textit{Phys.\ Rev.\ D} {\bf 96}, no. 5, 056012 (2017)}
  Addendum: \textcolor{red}{[\textit{Phys.\ Rev.\ D} {\bf 96}, no. 7, 079901 (2017)]}.
 % doi:10.1103/PhysRevD.96.079901, 10.1103/PhysRevD.96.056012
 % [arXiv:1701.08263 [hep-th]].
 
 %\cite{Boldizsar:2018akg}
\bibitem{Boldizsar:2018akg} 
  B.~Boldizsár, M.~I.~Nagy and M.~Csanád,
  %``Polarized Baryon Production in Heavy Ion Collisions: An Analytic Hydrodynamical Study,''
  \textcolor{red}{\textit{Universe} {\bf 5}, no. 5, 101 (2019)}.
  %doi:10.3390/universe5050101
 % [arXiv:1812.05587 [hep-ph]].
 
 %\cite{Liu:2019krs}
\bibitem{Liu:2019krs} 
  S.~Y.~F.~Liu, Y.~Sun and C.~M.~Ko,
  %``Spin polarizations in a covariant angular momentum conserved chiral transport model,''
  arXiv:1910.06774 [nucl-th].
  %%CITATION = ARXIV:1910.06774;%%
  %1 citations counted in INSPIRE as of 15 Jan 2020
  
  %\cite{Florkowski:2019voj}
\bibitem{Florkowski:2019voj} 
  W.~Florkowski \textit{et al.},
  %``Longitudinal spin polarization in a thermal model,''
  \textcolor{red}{\textit{Phys.\ Rev.\ C} {\bf 100}, no. 5, 054907 (2019)}.
  %doi:10.1103/PhysRevC.100.054907
  %[arXiv:1904.00002 [nucl-th]].
  %%CITATION = doi:10.1103/PhysRevC.100.054907;%%
  %9 citations counted in INSPIRE as of 15 Jan 2020

%\cite{Xie:2019jun}
\bibitem{Xie:2019jun} 
  Y.~Xie, D.~Wang and L.~P.~Csernai,
  %``Fluid Dynamics Study of the $\Lambda$ Polarization for Au+Au Collisions at $\sqrt{s_{NN}}=200$ GeV,''
  arXiv:1907.00773 [hep-ph].
  %%CITATION = ARXIV:1907.00773;%%
  %5 citations counted in INSPIRE as of 15 Jan 2020
  
  %\cite{Wu:2019eyi}
\bibitem{Wu:2019eyi} 
  H.~Z.~Wu \textit{et al.},
  %``Local spin polarization in high energy heavy ion collisions,''
  \textcolor{red}{\textit{Phys.\ Rev.\ Research.\ } {\bf 1}, 033058 (2019)}.
 % doi:10.1103/PhysRevResearch.1.033058
  %[arXiv:1906.09385 [nucl-th]].
  %%CITATION = doi:10.1103/PhysRevResearch.1.033058;%%
  %5 citations counted in INSPIRE as of 15 Jan 2020
  
  
  %\cite{Becattini:2019ntv}
\bibitem{Becattini:2019ntv} 
  F.~Becattini, G.~Cao and E.~Speranza,
  %``Polarization transfer in hyperon decays and its effect in relativistic nuclear collisions,''
  \textcolor{red}{\textit{Eur.\ Phys.\ J.\ C} {\bf 79}, no. 9, 741 (2019)}.
  %doi:10.1140/epjc/s10052-019-7213-6
  %[arXiv:1905.03123 [nucl-th]].
  %%CITATION = doi:10.1140/epjc/s10052-019-7213-6;%%
  %10 citations counted in INSPIRE as of 15 Jan 2020
  
  %\cite{Zhang:2019xya}
\bibitem{Zhang:2019xya} 
  J.~j.~Zhang \textit{et al.},
  %``A microscopic description for polarization in particle scatterings,''
  \textcolor{red}{\textit{Phys.\ Rev.\ C} {\bf 100}, no. 6, 064904 (2019)}.
  %doi:10.1103/PhysRevC.100.064904
 % [arXiv:1904.09152 [nucl-th]].
  %%CITATION = doi:10.1103/PhysRevC.100.064904;%%
  %8 citations counted in INSPIRE as of 15 Jan 2020
  
%\cite{Fukushima:2020qta}
\bibitem{Fukushima:2020qta} 
  K.~Fukushima and S.~Pu,
  %``Relativistic decomposition of the orbital and the spin angular momentum in chiral physics and Feynman's angular momentum paradox,''
  arXiv:2001.00359 [hep-ph].
  %%CITATION = ARXIV:2001.00359;%%
  
  %\cite{Florkowski:2019gio}
\bibitem{Florkowski:2019gio} 
  W.~Florkowski, A.~Kumar and R.~Ryblewski,
  %``Spin chemical potential for relativistic particles with spin 1/2,''
  arXiv:1907.09835 [nucl-th].
  %%CITATION = ARXIV:1907.09835;%%
  
  %\cite{Li:2019qkf}
\bibitem{Li:2019qkf} 
  S.~Li and H.~U.~Yee,
  %``Quantum Kinetic Theory of Spin Polarization of Massive Quarks in Perturbative QCD: Leading Log,''
  \textcolor{red}{\textit{Phys.\ Rev.\ D} {\bf 100}, no. 5, 056022 (2019)}.
 % doi:10.1103/PhysRevD.100.056022
 % [arXiv:1905.10463 [hep-ph]].
  %%CITATION = doi:10.1103/PhysRevD.100.056022;%%
  %1 citations counted in INSPIRE as of 15 Jan 2020
  
  %\cite{Hattori:2019ahi}
\bibitem{Hattori:2019ahi} 
  K.~Hattori, Y.~Hidaka and D.~L.~Yang,
  %``Axial Kinetic Theory and Spin Transport for Fermions with Arbitrary Mass,''
  \textcolor{red}{\textit{Phys.\ Rev.\ D} {\bf 100}, no. 9, 096011 (2019)}.
  % doi:10.1103/PhysRevD.100.096011
 % [arXiv:1903.01653 [hep-ph]].
  %%CITATION = doi:10.1103/PhysRevD.100.096011;%%
  %11 citations counted in INSPIRE as of 15 Jan 2020
 %\cite{Weickgenannt:2019dks}
\bibitem{Weickgenannt:2019dks} 
  N.~Weickgenannt, X.~L.~Sheng, E.~Speranza, Q.~Wang and D.~H.~Rischke,
  %``Kinetic theory for massive spin-1/2 particles from the Wigner-function formalism,''
  \textcolor{red}{\textit{Phys.\ Rev.\ D} {\bf 100}, no. 5, 056018 (2019)}.
 % doi:10.1103/PhysRevD.100.056018
  %[arXiv:1902.06513 [hep-ph]].
  %%CITATION = doi:10.1103/PhysRevD.100.056018;%%
  %14 citations counted in INSPIRE as of 15 Jan 2020

%\cite{Hattori:2019lfp}
\bibitem{Hattori:2019lfp} 
  K.~Hattori, M.~Hongo, X.~G.~Huang, M.~Matsuo and H.~Taya,
  %``Fate of spin polarization in a relativistic fluid: An entropy-current analysis,''
  \textcolor{red}{\textit{Phys.\ Lett.\ B} {\bf 795}, 100 (2019)}.
 % doi:10.1016/j.physletb.2019.05.040
 % [arXiv:1901.06615 [hep-th]].
  %%CITATION = doi:10.1016/j.physletb.2019.05.040;%%
  %13 citations counted in INSPIRE as of 15 Jan 2020
  
  %\cite{Deng:2020ygd}
\bibitem{Deng:2020ygd} 
  X.~G.~Deng \textit{et al.},
  %``Vorticity in low-energy heavy-ion collisions,''
  arXiv:2001.01371 [nucl-th].
  %%CITATION = ARXIV:2001.01371;%%
  
  %\cite{Ambrus:2019ayb}
\bibitem{Ambrus:2019ayb} 
  V.~E.~Ambrus,
  %``Helical massive fermions under rotation,''
  arXiv:1912.09977 [nucl-th].
  %%CITATION = ARXIV:1912.09977;%%
  %1 citations counted in INSPIRE as of 15 Jan 2020
  
  %\cite{Sheng:2019kmk}
\bibitem{Sheng:2019kmk} 
  X.~L.~Sheng, L.~Oliva and Q.~Wang,
  %``What can we learn from global spin alignment of $\phi$ meson in heavy-ion collisions?,''
  arXiv:1910.13684 [nucl-th].
  %%CITATION = ARXIV:1910.13684;%%
  
  %\cite{Ivanov:2019wzg}
\bibitem{Ivanov:2019wzg} 
  Y.~B.~Ivanov, V.~D.~Toneev and A.~A.~Soldatov,
  %``Vorticity and Particle Polarization in Relativistic Heavy-Ion Collisions,''
  arXiv:1910.01332 [nucl-th].
  %%CITATION = ARXIV:1910.01332;%%
  %2 citations counted in INSPIRE as of 15 Jan 2020
  
%\cite{Prokhorov:2019cik}
\bibitem{Prokhorov:2019cik} 
  G.~Y.~Prokhorov, O.~V.~Teryaev and V.~I.~Zakharov,
  %``Unruh effect for fermions from the Zubarev density operator,''
  \textcolor{red}{\textit{Phys.\ Rev.\ D} {\bf 99}, no. 7, 071901 (2019)}.
 % doi:10.1103/PhysRevD.99.071901
 % [arXiv:1903.09697 [hep-th]].
  %%CITATION = doi:10.1103/PhysRevD.99.071901;%%
  %7 citations counted in INSPIRE as of 15 Jan 2020
  
  %\cite{Prokhorov:2018bql}
\bibitem{Prokhorov:2018bql} 
  G.~Y.~Prokhorov, O.~V.~Teryaev and V.~I.~Zakharov,
  %``Effects of rotation and acceleration in the axial current: density operator vs Wigner function,''
  \textcolor{red}{\textit{JHEP} {\bf 1902}, 146 (2019)}.
 % doi:10.1007/JHEP02(2019)146
 % [arXiv:1807.03584 [hep-th]].
  %%CITATION = doi:10.1007/JHEP02(2019)146;%%
  %8 citations counted in INSPIRE as of 15 Jan 2020
  %\cite{Yang:2018lew}
\bibitem{Yang:2018lew} 
  D.~L.~Yang,
  %``Side-Jump Induced Spin-Orbit Interaction of Chiral Fluids from Kinetic Theory,''
  \textcolor{red}{\textit{Phys.\ Rev.\ D} {\bf 98}, no. 7, 076019 (2018)}.
%  doi:10.1103/PhysRevD.98.076019
 % [arXiv:1807.02395 [nucl-th]].
  %%CITATION = doi:10.1103/PhysRevD.98.076019;%%
  %8 citations counted in INSPIRE as of 15 Jan 2020


  

  
 %\cite{Becattini:2016gvu}
\bibitem{Becattini:2016gvu} 
  F.~Becattini \textit{et al.},
  %``Global hyperon polarization at local thermodynamic equilibrium with vorticity, magnetic field and feed-down,''
  \textcolor{red}{\textit{Phys.\ Rev.\ C} {\bf 95}, no. 5, 054902 (2017)}.
 % doi:10.1103/PhysRevC.95.054902
  %[arXiv:1610.02506 [nucl-th]].
  %%CITATION = doi:10.1103/PhysRevC.95.054902;%%
  %94 citations counted in INSPIRE as of 14 Jan 2020
  
  %\cite{Karpenko:2016jyx}
\bibitem{Karpenko:2016jyx} 
  I.~Karpenko and F.~Becattini,
  %``Study of $\Lambda $ polarization in relativistic nuclear collisions at $\sqrt{s_\mathrm {NN}}=7.7$ –200 GeV,''
  \textcolor{red}{\textit{Eur.\ Phys.\ J.\ C} {\bf 77}, no. 4, 213 (2017)}.
  %doi:10.1140/epjc/s10052-017-4765-1
 % [arXiv:1610.04717 [nucl-th]].
  %%CITATION = doi:10.1140/epjc/s10052-017-4765-1;%%
  %58 citations counted in INSPIRE as of 14 Jan 2020
  
  %\cite{Li:2017slc}
\bibitem{Li:2017slc} 
  H.~Li \textit{et al.},
  %``Global $\Lambda$ polarization in heavy-ion collisions from a transport model,''
  \textcolor{red}{\textit{Phys.\ Rev.\ C} {\bf 96}, no. 5, 054908 (2017)}.
 % doi:10.1103/PhysRevC.96.054908
 % [arXiv:1704.01507 [nucl-th]].
  %%CITATION = doi:10.1103/PhysRevC.96.054908;%%
  %49 citations counted in INSPIRE as of 14 Jan 2020

%\cite{Xie:2017upb}
\bibitem{Xie:2017upb} 
  Y.~Xie, D.~Wang and L.~P.~Csernai,
  %``Global Λ polarization in high energy collisions,''
  \textcolor{red}{\textit{Phys.\ Rev.\ C} {\bf 95}, no. 3, 031901 (2017)}.
  %doi:10.1103/PhysRevC.95.031901
 % [arXiv:1703.03770 [nucl-th]].
  %%CITATION = doi:10.1103/PhysRevC.95.031901;%%
  %40 citations counted in INSPIRE as of 14 Jan 2020
 
   

  
 %\cite{Becattini:2007sr}
\bibitem{Becattini:2007sr} 
  F.~Becattini, F.~Piccinini and J.~Rizzo,
  %``Angular momentum conservation in heavy ion collisions at very high energy,''
  \textcolor{red}{\textit{Phys.\ Rev.\ C} {\bf 77}, 024906 (2008)}.
 % doi:10.1103/PhysRevC.77.024906
%  [arXiv:0711.1253 [nucl-th]].
  %%CITATION = doi:10.1103/PhysRevC.77.024906;%%
  %141 citations counted in INSPIRE as of 14 Jan 2020
  
  

  %\cite{Florkowski:2017ruc}
\bibitem{Florkowski:2017ruc} 
  W.~Florkowski \textit{et al.},
  %``Relativistic fluid dynamics with spin,''
  \textcolor{red}{\textit{Phys.\ Rev.\ C} {\bf 97}, no. 4, 041901 (2018)}.
  %doi:10.1103/PhysRevC.97.041901
 % [arXiv:1705.00587 [nucl-th]].
 

  %\cite{Florkowski:2017dyn}
\bibitem{Florkowski:2017dyn} 
  W.~Florkowski \textit{et al.},
  %``Spin-dependent distribution functions for relativistic hydrodynamics of spin-1/2 particles,''
  \textcolor{red}{\textit{Phys.\ Rev.\ D} {\bf 97}, no. 11, 116017 (2018)}.
 % doi:10.1103/PhysRevD.97.116017
 % [arXiv:1712.07676 [nucl-th]].
  
    %\cite{Florkowski:2018fap}
\bibitem{Florkowski:2018fap} 
  W.~Florkowski, R.~Ryblewski and A.~Kumar,
  %``Relativistic hydrodynamics for spin-polarized fluids,''
  \textcolor{red}{\textit{Prog.\ Part.\ Nucl.\ Phys.\ } {\bf 108}, 103709 (2019)}.
  %doi:10.1016/j.ppnp.2019.07.001
  %[arXiv:1811.04409 [nucl-th]].
  
  
   %\cite{Florkowski:2018ahw}
\bibitem{Florkowski:2018ahw} 
  W.~Florkowski, A.~Kumar and R.~Ryblewski,
  %``Thermodynamic versus kinetic approach to polarization-vorticity coupling,''
  \textcolor{red}{\textit{Phys.\ Rev.\ C} {\bf 98}, no. 4, 044906 (2018)}.
  %doi:10.1103/PhysRevC.98.044906
  %[arXiv:1806.02616 [hep-ph]].
  
   %\cite{Florkowski:2019qdp}
\bibitem{Florkowski:2019qdp} 
  W.~Florkowski \textit{et al.},
  %``Spin polarization evolution in a boost invariant hydrodynamical background,''
  \textcolor{red}{\textit{Phys.\ Rev.\ C} {\bf 99}, no. 4, 044910 (2019)}.
 % doi:10.1103/PhysRevC.99.044910
 % [arXiv:1901.09655 [hep-ph]].
  
  %\cite{DeGroot:1980dk}
\bibitem{DeGroot:1980dk} 
  S.~R.~De Groot, W.~A.~Van Leeuwen, C.~G.~Van Weert,
  \textit{Relativistic Kinetic Theory, Principles and Applications},
  Amsterdam, North-Holland, 1980.
  
  
  
   %\cite{Bjorken:1982qr}
\bibitem{Bjorken:1982qr} 
  J.~D.~Bjorken,
  %``Highly Relativistic Nucleus-Nucleus Collisions: The Central Rapidity Region,''
  \textcolor{red}{\textit{Phys.\ Rev.\ D} {\bf 27}, 140 (1983)}.
 % doi:10.1103/PhysRevD.27.140
  
 
 
%   %\cite{Florkowski:2017olj}
% \bibitem{Florkowski:2017olj} 
%   W.~Florkowski, M.~P.~Heller and M.~Spalinski,
%   %``New theories of relativistic hydrodynamics in the LHC era,''
%   \textcolor{red}{\textit{Rept.\ Prog.\ Phys.\ }  {\bf 81}, no. 4, 046001 (2018)}.
%  % doi:10.1088/1361-6633/aaa091
%  % [arXiv:1707.02282 [hep-ph]].
 
  

  %\cite{Florkowski:2018myy}
%\bibitem{Florkowski:2018myy} 
 % W.~Florkowski, E.~Speranza and F.~Becattini,
  %``Perfect-fluid hydrodynamics with constant acceleration along the stream lines and spin polarization,''
  %Acta Phys.\ Polon.\ B {\bf 49}, 1409 (2018)
 % doi:10.5506/APhysPolB.49.1409
 % [arXiv:1803.11098 [nucl-th]].
  
  
  
%   %\cite{Montenegro:2017lvf}
% \bibitem{Montenegro:2017lvf} 
%   D.~Montenegro, L.~Tinti and G.~Torrieri,
%   %``Sound waves and vortices in a polarized relativistic fluid,''
%   Phys.\ Rev.\ D {\bf 96}, no. 7, 076016 (2017)
%  % doi:10.1103/PhysRevD.96.076016
%  % [arXiv:1703.03079 [hep-th]].
 
  
 
 
  
%   %\cite{Florkowski:2010zz}
% \bibitem{Florkowski:2010zz} 
%   W.~Florkowski,
%   %``Phenomenology of Ultra-Relativistic Heavy-Ion Collisions,''
%   Singapore, Singapore: World Scientific (2010) 416 p


% %\cite{Liu:2019krs}
% \bibitem{Liu:2019krs} 
%   S.~Y.~F.~Liu, Y.~Sun and C.~M.~Ko,
%   %``Spin polarizations in a covariant angular momentum conserved chiral transport model,''
%   arXiv:1910.06774 [nucl-th].
  
%   %\cite{Sun:2018bjl}
% \bibitem{Sun:2018bjl} 
%   Y.~Sun and C.~M.~Ko,
%   %``Azimuthal angle dependence of the longitudinal spin polarization in relativistic heavy ion collisions,''
%   \textcolor{red}{\textit{Phys.\ Rev.\ C} {\bf 99}, no. 1, 011903 (2019)}.
%   %doi:10.1103/PhysRevC.99.011903
%   %[arXiv:1810.10359 [nucl-th]].
  
  
  
%   %\cite{Voloshin:2004ha}
% \bibitem{Voloshin:2004ha} 
%   S.~A.~Voloshin,
%   %``Polarized secondary particles in unpolarized high energy hadron-hadron collisions?,''
%   nucl-th/0410089.

\end{thebibliography}
\end{document}